\UseRawInputEncoding

\documentclass[aps,prl,amsmath,amssymb,floatfix,reprint,citeautoscript,noeprint,superscriptaddress,twocolumn]{revtex4-2}
\usepackage{lineno}

\usepackage{dsfont}
\usepackage{lipsum} 
\usepackage{bibentry}
\usepackage[english]{babel}
\usepackage[dvipsnames]{xcolor}
\usepackage{graphicx}
\usepackage[caption=false]{subfig} 
\usepackage{amsmath,amssymb,bm}
\usepackage[version=3]{mhchem}
\usepackage{verbatim}
\usepackage{multirow}
\usepackage{dcolumn}
\usepackage{float}
\usepackage{nicefrac}
\usepackage{siunitx}
\usepackage{booktabs}
\usepackage{chemformula}
\usepackage{wrapfig}
\usepackage{enumitem}  
\usepackage{transparent}
\usepackage[colorlinks,allcolors=black,citecolor=blue,urlcolor=blue]{hyperref}
\usepackage{tcolorbox}
\usepackage{relsize}
\hypersetup{
    citecolor=Turquoise,
    colorlinks=true,
    linkcolor=black,    
    urlcolor=Turquoise,
    }
\usepackage{titlesec}

\titleformat{\section}
  {\normalfont\normalsize\bfseries} 
  {} 
  {0pt} 
  {} 
  [\vspace{-5pt}] 

\newcommand{\tbf}[1]{\textbf{#1}}

\newcommand{\mbf}[1]{\boldsymbol{\mathit{#1}}}
\newcommand{\mrm}[1]{\mathrm{#1}}
\newcommand{\mcl}[1]{\mathcal{#1}}
\newcommand{\tcr}[1]{\textcolor{black}{#1}}

\usepackage{sansmathfonts}
\usepackage[justification=centerlast, font={sf}, labelfont={bf}]{caption}

\usepackage{etoolbox}
\pretocmd{\caption}{\nolinenumbers}{}{}
\apptocmd{\caption}{\linenumbers}{}{}
\usepackage[left=1.8cm, right=1.35cm, top=1.8cm, bottom=1.8cm]{geometry}

\usepackage{pdfpages}

\makeatletter
\AtBeginDocument{\let\LS@rot\@undefined}
\makeatother
\begin{document}


\title{Dielectrocapillarity for exquisite control of fluids}

\author{Anna T. Bui}
\affiliation{Yusuf Hamied Department of Chemistry, University of
  Cambridge, Lensfield Road, Cambridge, CB2 1EW, United Kingdom}
\affiliation{Department of Chemistry, Durham University, South Road,
  Durham, DH1 3LE, United Kingdom}

\author{Stephen J. Cox}
\email{stephen.j.cox@durham.ac.uk}
\affiliation{Department of Chemistry, Durham University, South Road,
  Durham, DH1 3LE, United Kingdom}

\date{\today}

\begin{abstract}
Spatially varying electric fields are prevalent throughout nature, such as 
in nanoporous materials 
and biological membranes, and technology, e.g, patterned electrodes and van der Waals heterostructures. While uniform fields cause free ions to migrate,
for polar fluids they simply reorient the constituent molecules. In contrast, electric field gradients (EFGs) induce a dielectrophoretic force, 
offering fine control of polar fluids even in the absence of free charges. Despite their vast potential for optimizing fluid behavior,
EFGs remain largely unexplored
at the microscopic level due to the absence of a rigorous 
first-principles theory of electrostriction.  
By integrating state-of-the-art advances in liquid state theory and deep
learning, we reveal how EFGs modulate fluid structure and capillarity.  We demonstrate that
dielectrophoretic coupling enables tunable control over the
liquid--gas phase transition, capillary condensation, and fluid uptake
into porous media.  Our findings establish
``dielectrocapillarity''---the use of EFGs to manipulate confined
fluids---as a powerful mechanism for controlling volumetric capacity
in nanopores, holding immense potential
for energy
storage, 
selective gas separation, 
and tunable
hysteresis in neuromorphic nanofluidics. 
Furthermore, by linking nanoscale dielectrocapillarity to macroscopic 
dielectrowetting, we establish a foundation for field-controlled wetting and adsorption phenomena of polar fluids across length scales.

\end{abstract}

\maketitle

Nanoporous materials such as metal-organic frameworks \cite{Chen2020},
carbon-based supercapacitors \cite{Chmiola2006,Liu2024} and
nanofluidic devices \cite{Yang2020, Fumagalli2018,Bocquet2020,Noy2023}
rely on their ability to uptake and store fluids, in either the
gaseous or liquid state, which directly impacts the performance of
energy storage \cite{Salanne2016}, chemical separation
\cite{Sholl2016} and filtration technologies \cite{Gin2011}. \tcr{The
  physics of capillarity plays a fundamental role in determining fluid
  uptake in these systems. It is well established---initially at the
  macroscopic scale through the works of Young, Laplace, and Kelvin
  \cite{deGennes2003book, Fisher1981nature}, and later at the
  microscopic scale \cite{Evans1990nanopore, rowlinson2002book}---that
  adsorption depends not only on the \tcr{system's thermodynamic
    state}, but also on the confinement length and the
  substrate--fluid interaction. These factors are typically intrinsic
  material properties.  As a result,} extensive research into
enhancing adsorption in porous materials has focused on optimizing
these factors, e.g., by tuning porosity or chemical functionalization
\cite{Gu2019}. \tcr{However,} the potential for manipulation by
external means---using applied fields to control confined
fluids---remains relatively unexplored.
  
Electric fields offer a compelling mechanism to control the structure,
phase behavior \cite{Tsori2004} and interfacial properties
\cite{Hayes2003, McHale2011} of fluids.  While a uniform field exerts
a \tcr{direct} force only on free charges such as ions, non-uniform
fields with electric field gradients (EFGs) generate dielectrophoretic
forces on neutral polar molecules. \tcr{Dielectrowetting experiments have demonstrated that EFGs influence
  macroscopic contact angles, in a manner consistent with a modified
  Young's equation \cite{Mugele2005, McHale2011,McHale2013,
    Edwards2018}. Whether these effects translate to nanoscale
  capillarity, however, is unclear. Addressing this issue is
  important, as porous media and membranes are rarely uniform; surface
  heterogeneities \cite{Liu2024, Lee1994}, defects \cite{Dupuis2022}
  and curvature are natural sources of large EFGs. Such inhomogeneous
  fields can also be engineered using, e.g., patterned electrodes
  \cite{Hayes2003, McHale2011, Han2015}, atomic force microscopes
  \cite{Fumagalli2018}, or layered van der Waals heterostructures}
\cite{Yang2020, Castellanos-Gomez2022}.

\tcr{Such EFGs introduce new length scales that can be comparable to
the natural correlation lengths of the confined fluid, posing a
severe challenge for a comprehensive theoretical description.  With
their inherent microscopic resolution,} molecular dynamics simulations
have provided key insights into the structural response of fluids to
electric fields \cite{Merlet2012} and phase behavior under
confinement \cite{Kapil2022}.  However, computational limitations mean
that simulation studies typically fix the number of molecules in the
system, which introduces mechanical strains in an uncontrolled
fashion.  As a result, existing approaches---whether experimental,
computational, or theoretical---face significant limitations in
efficiently resolving both the microscopic restructuring and emergent
macroscopic reorganization of fluids in non-uniform electric fields.

Here, we investigate how EFGs 
on the molecular length scale
can be harnessed to manipulate mesoscopic fluid properties and phase
behavior.  Bringing together the latest advances in liquid state
theory, computer simulation, and machine learning, we develop a
multiscale framework to study electrostriction in polar
fluids---that is, their density response to applied electric
fields---within the grand canonical ensemble, 
representative of real conditions in which fluid molecules
can enter and leave a pore. We show that EFGs provide tunable
control over the liquid--gas phase transition and directly influence
adsorption capability by capillary condensation---we dub this new
phenomenon ``dielectrocapillarity.'' Given the critical role of
volumetric fluid uptake in nanoporous materials for energy
storage \cite{Chen2020}, gas separation \cite{Sholl2016}, and
filtration technologies \cite{Gin2011}, our findings establish
dielectrocapillarity as a promising avenue in fine-tuning and
optimization of such processes.  Furthermore, the ability
to regulate hysteresis introduces a new level of programmability
in nanofluidic systems, where EFG-driven control of adsorption and
desorption rates could offer external tunability akin to synaptic
plasticity in neuromorphic nanofluidic circuits \cite{Xiong2023,
Robin2021}.

\section*{Multiscale approach for electrostriction in fluids}

\begin{figure}[t]
  \includegraphics[width=0.95\linewidth]{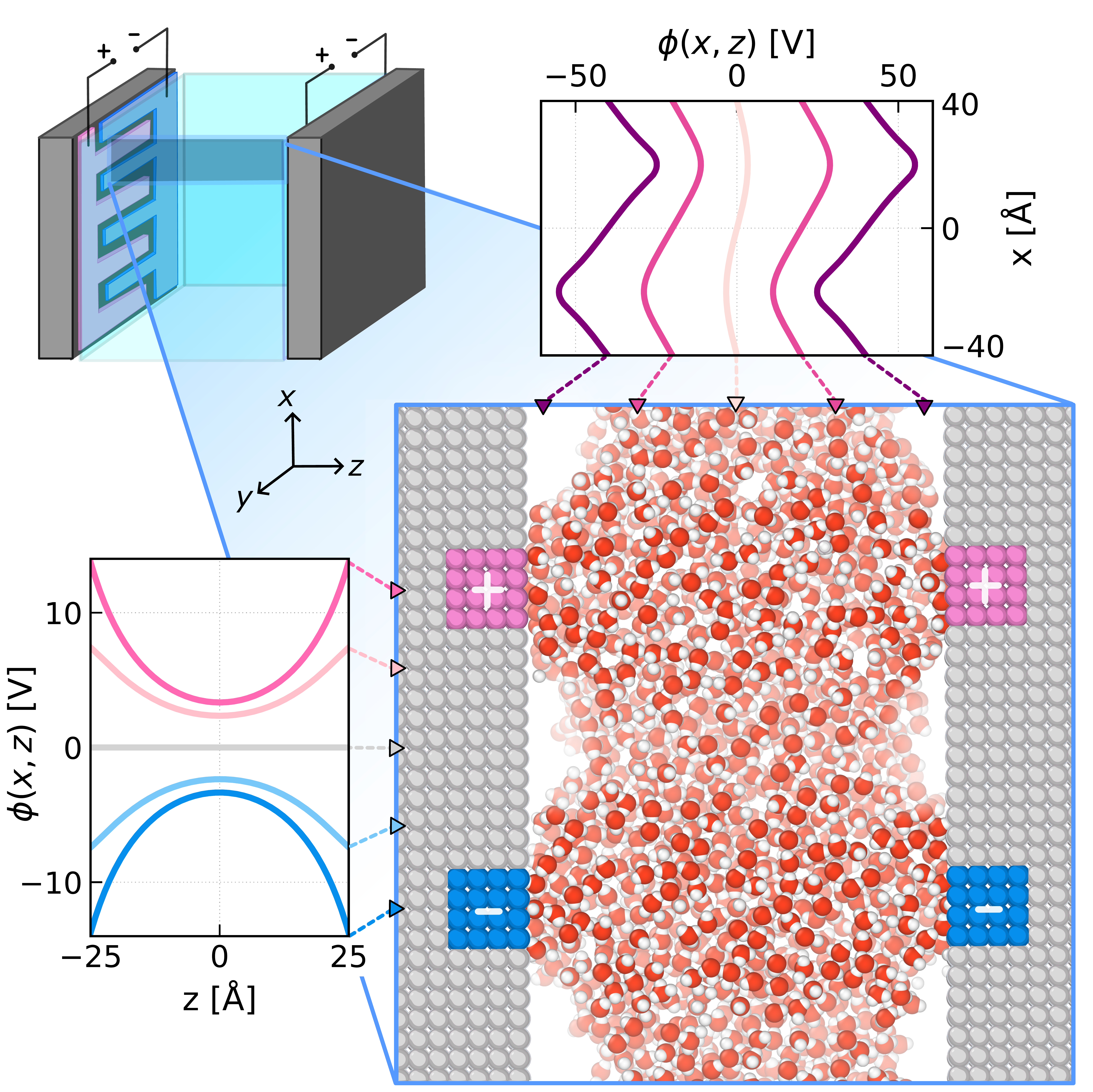}

  \caption{\tcr{\textbf{Inhomogeneous electric fields arising from
        interdigitated electrodes strongly influence water's wetting
        behavior.} Snapshot of an SPC/E water simulation in a
      hydrophobic slit with alternating positive and negative
      electrode patches. Either holding the electrodes at a 10\,V
      potential difference or attributing a fixed charge of $\pm
      0.05\,e/\mathrm{atom}$ causes the fluid to exhibit enhanced
      wetting at the walls, accompanied by strong lateral density
      oscillations. Cross-sections of the electrostatic potential in
      the constant charge setup are shown parallel to the surface (top
      right) and normal to the surface (bottom left).}}
\label{fig0}
\end{figure}

\tcr{Although molecular simulations cannot fully capture the influence of
EFGs on dielectric fluids, they offer vivid qualitative insights into
the underlying physics and illustrate how EFGs may arise within
nanoscale devices, as illustrated in Fig.~\ref{fig0}. Here, we present
a simulation snapshot of water at 300\,K confined between two
hydrophobic substrates that have been patterned with alternating
stripes of positive and negative charge---a set up that provides a
caricature of an interdigitated electrode. As can be clearly seen in
Fig.~\ref{fig0}, charging this device induces local wetting near the
charged stripes, with pronounced density variations along the
direction parallel to the surface ($x$).}

\tcr{These features arise from the complex interplay between the fluid's
charge and number densities, and their collective response to EFGs,
arising from the inhomogeneous electrostatic potential $\phi(x,z)$,
where $z$ is along the surface normal. It is instructive to consider
the form of $\phi$ far from either surface; here, it will resemble the
linear superposition of the asymptotic limit of the two substrates
considered independently,
\begin{equation}
\begin{split}
    \phi_{\rm single}(x,z)  &\sim \phi_0 \sin\!\left(\frac{2\pi x}{L_x}\right)\exp\!\left(\frac{-2\pi |z-z_{\rm s}|}{L_x}\right),
\end{split}
\end{equation}
where $z_{\rm s}$ indicates the plane of the substrate's outermost
atoms. We therefore observe that inhomogeneity of $\phi$ is
characterized by a sinusoidal oscillation of period $L_x$ along $x$, and
an exponential decay along from the surface along $z$. This asymptotic
analysis captures the essential behavior of $\phi(x,z)$ computed
explicitly from the potential energy of a test charge, as seen in
Fig.~\ref{fig0}.}

\tcr{The pronounced wetting behavior observed with simulation in
Fig.~\ref{fig0} strongly suggests that such EFGs will influence
capillarity of the polar fluid; that is, the amount of fluid adsorbed
at constant chemical potential. Addressing this issue, however,
demands an accurate and efficient framework for determining structure,
thermodynamics, and phase behavior in an open system---this lies
beyond the practical limits of present day molecular
simulations. Instead, we turn to classical density functional theory
(cDFT), an exact statistical mechanical framework for inhomogeneous
fluids.}

\tcr{Within cDFT,} the equilibrium structure and thermodynamics of a fluid
can be determined from first principles by its excess intrinsic
Helmholtz free energy functional
$\mcl{F}_{\mrm{intr}}^{\mrm{ex}}([\rho],T)$, where $\rho(\mbf{r})$ is
the average inhomogeneous density of the fluid and $T$ is the
temperature. \tcr{This central result, established in
Ref.~\cite{Evans1979}, places cDFT as the liquid-state generalization}
of its celebrated electronic structure
counterpart \cite{Kohn1999}---as a modern theory for inhomogeneous
fluids. cDFT is naturally formulated in the grand canonical ensemble,
where the chemical potential acts as the control variable governing
particle exchange in confined systems. The chemical potential, $\mu$,
maps directly onto the relative humidity or vapor pressure for gases,
and chemical activity for liquids.

In practice, the exact form of
$\mcl{F}_{\mrm{intr}}^{\mrm{ex}}([\rho],T)$ is generally
unknown. Instead of relying on traditional approximations, we leverage
state-of-the-art data-driven methodologies to supervise-learn
functional mappings directly from quasi-exact reference data from
grand canonical Monte Carlo simulations \cite{Sammuller2023,
  Sammuller2025, bui2024learningclassicaldensityfunctionals}.
\tcr{This machine-learned cDFT framework has already been successfully
  applied to liquid--gas coexistence \cite{Sammuller2025},
  liquid--liquid phase separation \cite{Robitschko2025}, and the
  electric double layer
  \cite{bui2024learningclassicaldensityfunctionals}.}  Going beyond
established deep-learning approaches to cDFT, we capture
electrostriction arising from the coupling between mass and charge
density of the fluid by explicitly learning the ``hyperfunctional''
$\mcl{F}_{\mrm{intr}}^{\mrm{ex}}([\rho,\beta\phi],T)$ where
$\phi(\mbf{r})$ is the inhomogeneous electrostatic potential and
$\beta=1/(k_{\mrm{B}}T)$ with $k_{\mrm{B}}$ the Boltzmann constant, as
recently introduced in Ref.~\cite{bui2025hyperlmft}. \tcr{A practical
  limitation of this cDFT approach is that, at present, only
  inhomogeneities with planar symmetry can be investigated
  directly. As a result, the neural-network representation of this
  functional was trained on simulation data generated under random
  planar electrostatic potentials $\phi(z)$.} \tcr{Nonetheless, we
  demonstrate below that calculations in which $\phi$ only varies
  along a single cartesian direction provide general insight into the
  influence of EFGs on wetting and capillarity.}  Details of the
practical implementation of the theory are given in the Methods
section.

The resulting cDFT framework is not only efficiently computable on
standard hardware but also unparalleled in its ability to
simultaneously capture microscopic fluid structure and mesoscopic
phase behavior under arbitrary non-uniform electric fields. Unlike
atomistic simulations, which are computationally prohibitive for such
a broad exploration, our method achieves orders-of-magnitude speedup,
completing each calculation in less than a minute without sacrificing
quantitative accuracy.  \tcr{Importantly, this computational
  efficiency is realized after a one-time training stage: once the
  functional has been learned, it can be used to perform thousands of
  free-energy calculations at negligible cost. This amortized
  advantage is what enables us to map complete phase diagrams,
  adsorption isotherms, and metastable fluid branches---tasks that
  would require enhanced sampling and thermodynamic integration if
  carried out by molecular simulation alone.}  This efficiency enables
an unprecedented, highly accurate mapping of a polar fluid's response
to EFGs of varying strengths and wavelengths across different
thermodynamic conditions, from supercritical to subcritical regimes,
spanning both bulk and confined environments. With this powerful tool
in hand, we now uncover emergent electrostrictive phenomena that arise
from the complex interplay of thermodynamics, confinement, and
response to EFGs.

\begin{figure*}[t]
  \includegraphics[width=0.9\linewidth]{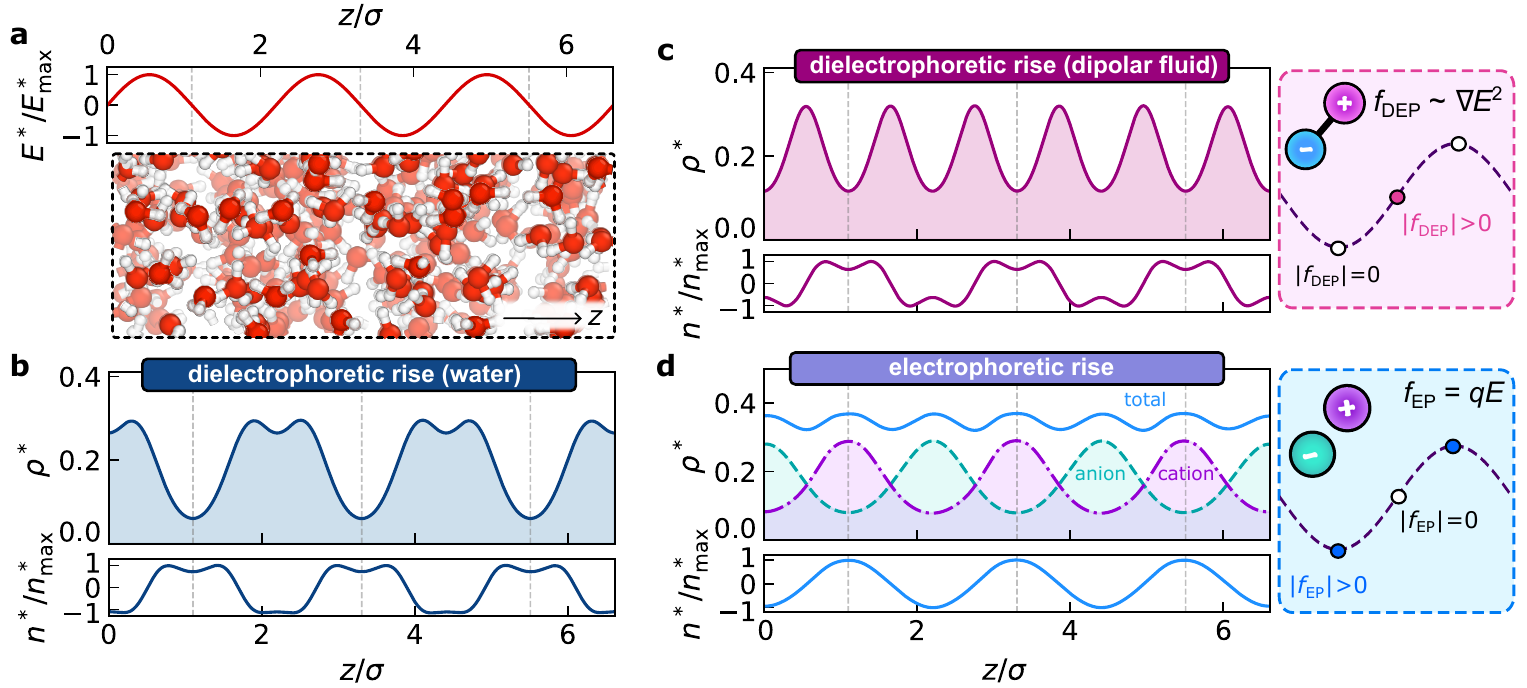}

  \caption{\textbf{Reorganization of fluids under non-uniform electric
  fields.} An applied electric field, $E^*(z) = E^*_{\rm
  max}\sin(2\pi z/\lambda)$, shown in \textbf{(a)}, induces pronounced
  density variations in bulk supercritical water, as can clearly be
  seen from a snapshot of a molecular dynamics
  simulation. \textbf{(b)} Results from cDFT for the number
  ($\rho^{*}(z)$, top) and charge ($n^{*}(z)$, bottom) densities
  capture this behavior. It can clearly be seen that number density is
  locally depleted where the $|\nabla E|$ is large, and locally
  enhanced where $|\nabla E|$ is small. The same qualitative behavior
  is seen in \textbf{(c)} for a supercritical dipolar fluid, except
  that its response is symmetric, in contrast to water where local
  depletion depends upon the sign of $\nabla E$. \textbf{(d)} In
  contrast to both water and the dipolar fluid, an electrolyte is
  locally depleted in regions of low field strength due to
  electrophoretic forces (purple and green lines show cation and anion
  density, respectively, while the blue line shows the total
  density). Reduced units are described in the Methods section.}  
\label{fig1}
\end{figure*}

To \tcr{systematically} investigate how EFGs influence fluids, we
primarily consider a minimal molecular model that incorporates
soft-core repulsion, van der Waals attraction, and long-range dipolar
interactions. The advantage of using such a simple molecular model is
that we can exhaustively explore a broad range of thermodynamic
conditions, while potentially uncovering common behaviors among polar
fluids, from molecular liquids to colloidal systems. In addition, we
also investigate a commonly used simple point charge model for water
(SPC/E \cite{Berendsen1987}) that explicitly incorporates
hydrogen-bonding, under thermodynamic conditions close to its critical
point.  Where comparison with ionic fluids is made, we will also show
results for a prototypical model comprising oppositely charged hard
spheres
\cite{bui2024learningclassicaldensityfunctionals}; in
this case we use a straightforward generalization of cDFT to
multicomponent systems.

\section*{Dielectrophoretic coupling with EFGs
under non-uniform electric fields}

\begin{figure*}[t]
  \includegraphics[width=0.9\linewidth]{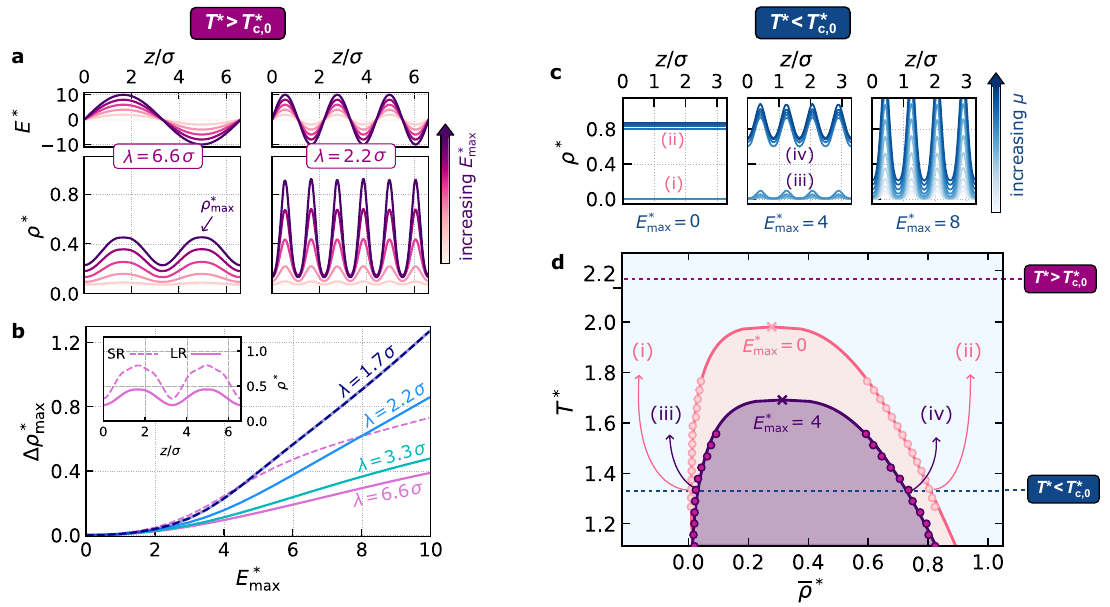}

  \caption{\textbf{Controlling liquid--vapor equilibrium with EFGs.}
    (\textbf{a}) At $T^* > T^*_{\rm c,0}$ where $T^*_{\rm c,0} \approx
    1.97$, increasing EFGs by independently varying $\lambda$ and
    $E_{\rm max}$ amplifies dielectrophoretic rise. (\textbf{b}) The
    local electrostrictive response of the dipolar fluid, as measured
    by the rise in the maximum density peak relative to zero field
    $\Delta\rho^*_{\rm max}$, is highly non-linear. Solid and dashed
    lines show the response of systems with dipolar interactions that
    are long-ranged (LR), e.g., polar molecules, and screened
    short-ranged (SR), e.g., colloids, respectively.  The effect of LR
    interactions becomes pronounced for $\lambda \gg \sigma$. The SR
    fluid is a nearly identical dipolar fluid, but whose Coulomb
    potential is replaced by $\mrm{erfc}(\kappa r)/r$ where
    $\kappa^{-1}=1.5\,\sigma$. (\textbf{c}) At an isotherm where $T^*
    < T^*_{\rm c,0}$, stable solutions for the density $\rho^\ast(z)$
    under a sinusoidal electric field with $\lambda/\sigma = 1.7$, are
    shown for different values of the chemical potential. These
    results are used to investigate liquid--vapor coexistence.
    (\textbf{d}) Results in light pink show the binodal of the dipolar
    fluid in the absence of an electric field.  At $E^*_{\rm max} =
    4$, $T^*_{\rm c}$ shifts to a lower temperature, as seen in the
    binodal in dark purple. Solid symbols show results obtained from
    the multiscale cDFT approach, while crosses indicate estimates of
    $T^*_{\rm c}$ using the law of rectilinear diameters and critical
    exponents \cite{rowlinson2002book}. Solid lines serve as a guide
    to the eye.}
\label{fig2}
\end{figure*}

\tcr{While EFGs will be most pronounced near the surfaces that
  generate them, Fig.~\ref{fig0} demonstrates that they may persist
  relatively far from the interface. In the specific case considered
  in Fig.~\ref{fig0}, midway between the substrates we observe a
  sinusoidal electrostatic potential along the $x$ direction. This
  motivates us to understand the direct influence of such sinusoidal
  potentials on the fluid, without explicitly considering
  interfaces. In fact, for $L_x$ larger than a few molecular
  diameters, we can define bulk response if we consider averages over
  a thin slice of thickness $\Delta z \ll L_x$ \tcr{(Fig.~S16)}. 
  Although such a clean separation of bulk and interfacial response
  becomes challenging as $L_x$ approaches molecular length scales, the
  full potential (as opposed to its asymptotic form) comprises modes
  of decreasing wavelength---it is therefore instructive to understand
  bulk-like response across a broad range of wavelengths.} 

While the behavior of fluids under uniform electric fields has been
extensively studied
\cite{Debye1965, Stevens1995, Zhang2020, Cassone2024},
non-uniform electric fields remain comparatively less
explored. Consequently, the effects of EFGs on fluids are less well
understood, even \tcr{in bulk}. As a starting point, we characterize
the bulk response of water, the simple polar fluid, and the
electrolyte, all under supercritical conditions, when subjected to a
sinusoidal electric field, $E^*(z) = E^*_{\rm max}\sin(2\pi
z/\lambda)$, as shown in Fig.~\ref{fig1}a. Quantities labeled with an
asterisk are expressed in reduced units, defined in the Methods
section.

In the case of water, Fig.~\ref{fig1}b, we see that the applied
electric field induces a significant structural reorganization along
the field direction; its average density profile $\rho^*(z)$ is
locally depleted in regions of weaker field strength, while molecular
reorientation leads to an inhomogeneous average charge density
distribution $n^\ast(z)$. As water is overall neutral and therefore
experiences no net electrophoretic force, the observed local
reorganization arises instead from dielectrophoretic forces,
$f_{\mrm{DEP}}\sim \nabla E^2$ \cite{Pohl1978,LandauLifshitzBook},
which push the fluid towards regions of higher electric field
strength---an effect termed ``dielectrophoretic rise.'' This
dielectrophoretic force is the same that drives dielectrophoresis,
which is widely exploited to manipulate biological
cells \cite{Pohl1966} and colloids \cite{Gangwal2008}, but its role in
molecular fluids has received little attention. While water provides
an important example of a polar fluid, the observed effects are by no
means specific to aqueous systems. As seen in Fig.~\ref{fig1}c, the
overall picture is the same for the simple polar fluid, aside from the
fact that it exhibits a symmetric response, whereas for water,
depletion is stronger in regions where $\nabla E < 0$ than where
$\nabla E > 0$ due to the inherent charge asymmetry of the water
molecule.

In contrast, applying a sinusoidal field to an electrolyte induces
``electrophoretic rise'' in which the fluid migrates toward regions of
lower electric field strength (Fig.~\ref{fig1}d). This is a result of
the electrophoretic force $f_{\mrm{EP}}=qE$, where $E$ is the local
field strength and $q$ is the ionic charge, causing the anions and
cations to reorganize, with peaks in their density profiles out of
phase due to their opposing charges. In this way, polar fluids and
ionic fluids display electromechanical responses that are
fundamentally distinct from each other.

Crucially, dielectrophoretic coupling depends on the EFGs as well
as the absolute field strength, and therefore offers greater control
over the fluid's response. To illustrate this, we present in
Fig.~\ref{fig2}a how dielectrophoretic rise can be amplified by
controlling the applied field. Owing to their qualitatively
similar behavior, in the remainder of the article we focus on the
dipolar fluid, with results for water given in the SI, Figs.~S13 and
S14. As the wavelength of the sinusoidal field decreases, local mass
accumulation becomes more pronounced, reflecting the system's response
to larger local EFGs, even as the maximum field amplitude remains
unchanged.  As a function of field strength, dielectrophoretic rise
exhibits strong non-linearity, as evident from the change in maximum
local density
$\Delta\rho^*_{\mrm{max}}=\rho^*_{\mrm{max}}-\rho^*_{\mrm{max},0}$
from zero field shown in Fig.~\ref{fig2}b, which highlights the
complexity and collectiveness of electrostrictive response in fluids.

Dielectrophoretic response is not solely determined by the magnitude
of the applied EFGs but also by the fluid's intermolecular
interactions.  This distinction is of practical importance when
considering colloidal systems in which effective interactions can be
tuned. For example, in electrolyte solutions, zwitterionic Janus
particles have diameters that far exceed the electrostatic screening
length, making their dipolar interactions inherently short-ranged
\cite{Hong2006}. To explore the effect on colloidal fluids, we
consider a nearly identical model fluid whose dipolar interactions are
screened, decaying on a length scale comparable to the molecular
diameter $\sigma$.  When the wavelength, $\lambda$, of the electric
field is comparable to the particle size, $\lambda \approx \sigma$,
both fluids exhibit identical dielectrophoretic response, shown by the
dashed blue line in Fig.~\ref{fig2}b. This result reflects that, for
wavelengths on the molecular scale, response is dominated by local
reordering of individual particles. However, for $\lambda\gg\sigma$,
behaviors differ significantly; for systems with long-ranged
interactions, molecular dipoles collectively reorient to screen the
applied field, weakening its effect over extended distances.  In
contrast, the short-ranged colloidal system lacks this screening,
leading to a much stronger response (dashed purple lines,
Fig.~\ref{fig2}b and inset). Such an \tcr{equilibrium} effect
\tcr{could} be leveraged for programmable directed self-assembly
\cite{McMullen2022}, where EFGs in combination with tunable Janus
particle surfaces \cite{Walther2013} \tcr{may} provide a powerful tool
for tailoring the assembly of extended structures with
dielectrophoretic forces, \tcr{with additional tunability arising from
  the solvent and ionic strength.} 

\section*{Fine tuning liquid-vapor coexistence with EFGs}

The results so far demonstrate that EFGs cause local
reorganization of a supercritical dielectric fluid into regions of low
and high density. A natural question then arises: how do EFGs
influence the phase behavior of such a single-component fluid?
Understanding this fundamental issue will be of central importance to
the optimal design of devices with switchable functionality, in a
similar spirit to the study of electric-field-induced phase
transformations in solid-state materials \cite{Lu2017, Zhang2019}.

In Fig.~\ref{fig2}c, we show the density profiles of the
fluid along the direction of an external sinusoidal electric
field with $\lambda/\sigma=1.7$, at $T^*<T^*_{\rm c,0}$, where
$T^*_{\rm c,0}$ is the critical temperature in the absence of an
external field. Results are shown for different chemical potentials.
At zero field, the stable solutions separate into homogeneous
vapor and liquid states at low and high chemical potential,
respectively. As the field strength increases, stronger EFGs (in
absolute terms) not only give rise to dielectrophoretic rise but also
destabilize liquid--vapor phase separation.  Consequently, within the
coexistence region, the vapor phase becomes denser, or contracts,
while the liquid phase expands. For sufficiently large
EFGs ($E^*_{\mrm{max}}=8$), the fluid undergoes a transition to
a single-phase supercritical fluid.
By locating vapor and liquid solutions with equal grand potentials,
in Fig.~\ref{fig2}d we map out the liquid--vapor binodal curve for the dipolar fluid, both
without an external field and under this sinusoidal field with
$E^*_{\mrm{max}} = 4$.  Strikingly, we find
that the critical temperature $T_{\mrm{c}}$ shifts downward under the
non-uniform field. To our knowledge, this marks the first report of a
shift in $T_{\mrm{c}}$ for a single-component fluid induced by an
electric field that varies on the microscopic length scale. Notably,
unlike for uniform fields \cite{Hegseth2004, Moore2015, Maerzke2010},
while the binodal line still represents liquid--vapor equilibrium, the
density within each phase is no longer spatially uniform due to
dielectrophoretic rise. In this case, the bulk density $\bar{\rho}$
represents the fluid's density averaged over a volume large compared
to $\sigma^3$.
For water, we observe similar behavior, with a downward shift in
$T_{\mrm{c}}$ of approx. $50\,\mathrm{K}$ at $E_{\rm
max}=0.4\,\mathrm{V\,\AA^{-1}}$ (Fig.~S14). These observations
highlight the exquisite level of control that one can exert over
dielectric fluids with EFGs.

Here, we have tuned the EFGs by varying $E^*_{\rm max}$ for fixed
$\lambda$. We could also have varied $\lambda$ for fixed $E^*_{\rm
  max}$, which provides additional control over the phase behavior
(Fig.~S11)\tcr{; importantly, these results demonstrate that the
  effects we have reported remain at larger wavelengths.} Since such
phase transitions are fully reversible, they will be particularly
relevant for functional nanofluidic and nano-electromechanical devices
where dynamic and reconfigurable phase control is advantageous
\cite{Fumagalli2018,Yang2020}. The act of confinement by itself
already leads to fluid behavior that can differ substantially from
bulk. EFGs represent an additional powerful tool for tailoring the
properties of fluids for the purposes of device design.

\section*{Controlling adsorption into porous media through dielectrocapillary phenomena}

\begin{figure}[t]
  \includegraphics[width=0.91\linewidth]{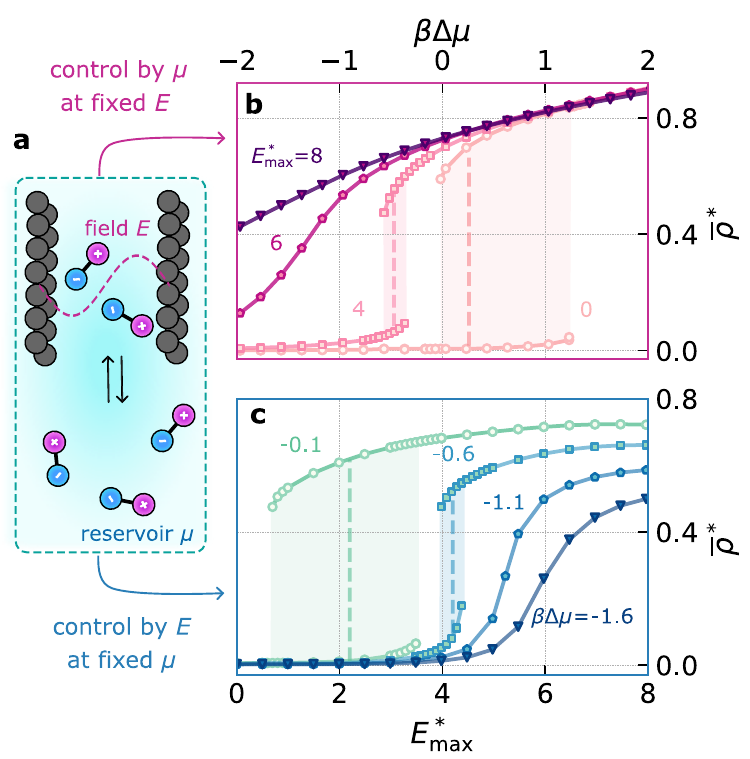} \caption{\textbf{Control
  of fluid uptake by dielectrocapillarity.  }  (\textbf{a}) Schematic
  of a fluid in a slit pore, with $H/\sigma = 6.6$, in equilibrium
  with a reservoir at chemical potential $\mu=\mu_{\rm
  co,0}+\Delta\mu$. (\textbf{b}) Adsorption/desorption isotherms at
  $T^*/T^*_{\rm c,0} = 0.68$ obtained by varying $\mu$ for different
  $E_{\rm max}$ at fixed $\lambda/\sigma = 1.7$. Larger $E_{\rm max}$
  promotes adsorption, while simultaneously decreasing hysteresis. For
  large enough $E^*_{\rm max}$, the transition becomes
  continuous. (\textbf{c}) Adsorption/desorption isotherms at
  $T^*/T^*_{\rm c,0} = 0.68$ obtained by varying $E_{\rm max}$ at
  fixed $\lambda/\sigma = 1.7$ for different $\Delta\mu$. Changing
  $E_{\rm max}$ can switch the pore between filled and empty
  states.  The vertical dashed lines indicate the equilibrium
  transition, i.e., where both adsorbed ``liquid'' and ``gas'' states
  are stable.}
\label{fig3}
\end{figure}

In a slit pore, uptake of a fluid can be monitored by isothermal
adsorption.  For fluids below their critical point, of which liquid
water at room temperature is an important example, adsorption is
governed by capillary effects, such that condensation can occur at
chemical potentials below saturation. This phenomenon, known as
capillary condensation, underlies the filling of nanochannels
\cite{Yang2020}, in which fluid uptake is controlled by adjusting the
relative humidity of the environment. Such an experimental setup
describes an equilibrium between the nanochannel and a reservoir
(Fig.~\ref{fig3}a), and maps directly on to our theoretical framework,
which is formulated in the grand canonical ensemble. In the absence of
an applied external field, \tcr{it is well-established that}
capillarity is controlled by the chemical potential of the reservoir,
\tcr{the length scale of confinement, the substrate--fluid interaction and temperature}
\cite{Evans1990nanopore}. With EFGs, we introduce an additional
experimental handle by which to control fluid adsorption behavior; we
call this new phenomenon ``dielectrocapillarity.''

We investigate a liquid at $T^*/T^*_{\rm c,0} = 0.68$ confined to a
solvophobic slit comprising two repulsive walls separated by a
distance $H = 6.6\,\sigma$. In Fig.~\ref{fig3}b, we show the fluid
uptake into the slit as a function of the chemical potential,
referenced to its coexistence value at zero field,
$\Delta\mu=\mu-\mu_{\rm co,0}$, typical of an adsorption/desorption
isotherm measurement.  In the absence of an electric field, the
transition is discontinuous, exhibiting a hysteresis loop---this is a
well-established hallmark of capillary condensation in nanopores and
mesopores, observed across experiments, simulations, and theory
\cite{Panagiotopoulos1987, Evans1990nanopore, Valiullin2006,
  Horikawa2011}.  Such hysteresis arises from the metastability of the
vapor during condensation and liquid during evaporation.

\tcr{To gain insight into the influence of EFGs, we perform a ``computational
experiment'' in which we apply a sinusoidal electric field across the
slit. While such a set up does not correspond to an EFG established by
the substrate walls themselves (see, e.g., Fig.~\ref{fig0}), it does
allow us to assess the effects of a particular mode. The impact is
twofold:} (1) condensation shifts to more negative $\Delta\mu$, i.e.,
the slit can be filled at even lower humidity; and (2) hysteresis is
reduced. Keeping $\lambda/\sigma = 1.7$ fixed, for sufficiently high
field strengths, hysteresis disappears entirely. In other words, we
have changed the nature of the transition from first-order to
continuous. This behavior directly results from the dielectrophoretic
coupling that drives the fluid toward its supercritical state
(Fig.~\ref{fig2}). Important for our fundamental understanding of
fluids under confinement, this result demonstrates that EFGs not only
shift the bulk critical temperature, but also influence the capillary
critical temperature.


Non-uniform electric fields clearly offer an additional lever for
controlling capillary filling.  As illustrated in Fig.~\ref{fig3}c,
adsorption/desorption can be actively controlled by switching the
field on or off, offering a precise, tunable means of regulating fluid
uptake. Such an ability to tailor hysteresis introduces a new level of
programmability in nanofluidic systems, where EFGs
can potentially serve as an external control parameter for
dynamically altering adsorption and desorption rates. Notably, tunable
hysteresis in capillary condensation can also serve as a memory
mechanism in neuromorphic nanofluidic circuits
\cite{Xiong2023, Robin2021}, where phase transitions encode
state-dependent responses akin to synaptic plasticity.

\section*{Connecting to dielectrowetting experiments}

\begin{figure}[t]
  \includegraphics[width=0.91\linewidth]{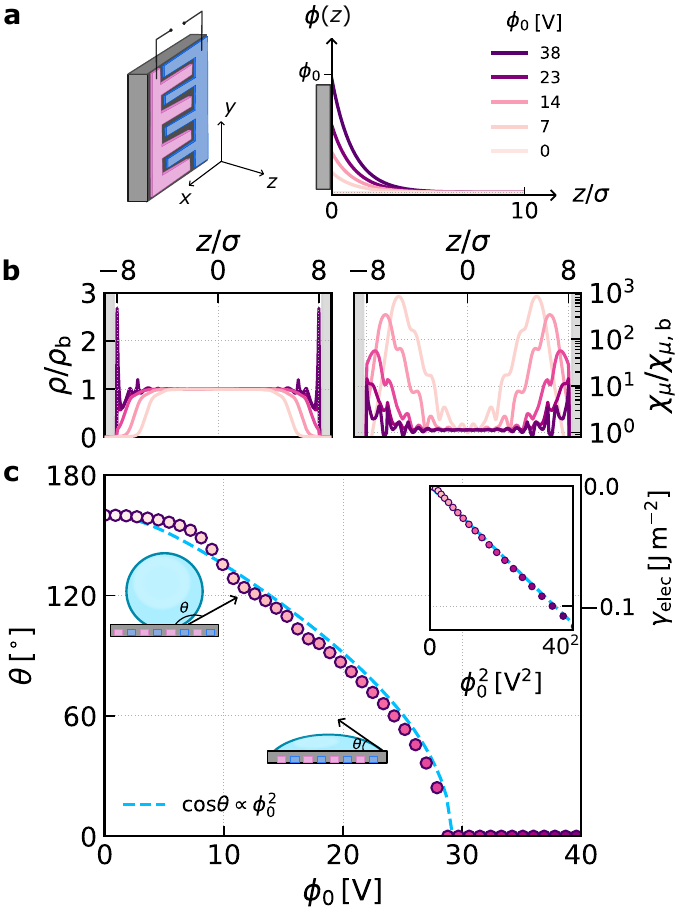}
  \caption{\textbf{Connecting to
  dielectrowetting experiments.} The electrostatic
  potential from interdigitated electrodes, shown schematically
  in \tbf{(a)}, decays exponentially. Applying this potential
  symmetrically from both confining walls in a slit geometry with
  $H\approx 16\,\sigma$ enhances wetting of the solid-liquid
  interface, as can be seen in the density profiles (left) and changes
  in local compressibility (right) in \textbf{(b)} (both quantities
  are normalized by their bulk values).} (\textbf{c}) Electrostatic
  free energy per unit area from cDFT (inset) exhibits a
  quadratic dependence on $\phi_{0}$, enabling reconstruction of the
  contact angle via Eq.~\ref{Eq:dielectrowetting} (assuming
  $\theta_0=160^\circ$ and using
  $\gamma_{\mrm{lv}}=0.025\mrm{J\,m^{-2}}$ computed from direct
  coexistence simulation), in direct analogy with dielectrowetting
  experiments \cite{McHale2011, McHale2013}.
\label{fig4}
\end{figure}

We have introduced dielectrocapillarity as an additional mechanism of
controlling capillary phenomena, complementing electrocapillarity
\cite{Grahame1947} where electrolytes respond to applied
potentials. On the macroscopic scale, these effects manifest in
electrowetting \cite{Mugele2005} and dielectrowetting
\cite{McHale2011, McHale2013}, where the contact angle of a droplet
can be tuned with applied potentials.  \tcr{Our nanoscale simulations
  in Fig.~\ref{fig0} show that a similar phenomenology emerges under
  confinement: wetting is strongly enhanced directly over the
  electrode patches, while the electrostatic potential generated by
  the interdigitated electrodes decays into the slit. Following
  previous experimental work \cite{McHale2011, McHale2013}, this
  motivates us employ a simple planar electrostatic potential,
  $\phi(z) =\phi_{0} \exp(-2\pi z/L_x)$ to probe wetting at the
  nanoscale. Similar to our arguments above, results obtained from
  such a potential can be considered to report on average behavior in
  a slice of thickness $\Delta x \ll L_x$ \tcr{(Fig.~S16)}. }

Within a macroscopic model, the resulting contact angle can be
described by a modified Young's law \cite{McHale2011, McHale2013},
\begin{equation}
    \cos \theta (\phi_{0}) =  \cos \theta_0 + \frac{\alpha}{\gamma_{\mrm{lv}}} \phi_{0}^2,
    \label{Eq:dielectrowetting}
\end{equation}
where $\theta_0$ is the Young's contact angle at zero field,
$\gamma_{\mathrm{lv}}$ the liquid--vapor surface tension, and $\alpha$
a material parameter related to the dielectric \tcr{permittivity} of
the liquid.  This relationship assumes that the electrostatic energy
stored in the liquid droplet is well-described by dielectric continuum
theory, and that \tcr{$L_x$} is sufficiently \tcr{small} so that any
changes in energy due to the electric field are effectively localized
to the solid-liquid interface. Under these assumptions, the
electrostatic free energy per unit area obeys the simple quadratic
scaling $\gamma_{\mathrm{elec}}=-\alpha \phi_{0}^2$.

Our multiscale framework provides a microscopic perspective on this
phenomenology, and allows us to test whether, at the nanoscale, EFGs
indeed enhance the wetting of dielectric liquids and the extent to
which the scaling prescribed by Eq.~\ref{Eq:dielectrowetting} holds.
To this end, \tcr{we applied $\phi(z)$ with $L_x\approx 7 \,\sigma$}
to the confined dipolar model, symmetrically from both walls of a
solvophobic slit. While not trained on such electrostatic potentials,
as can be seen in Fig.~\ref{fig4}b, the neural functional extrapolates
well, yielding physically plausible results, aside from some
oscillations at high wetting that are likely minor artifacts.  As
$\phi_0$ increases, so too does the contact density at the wall,
verifying that enhanced wetting occurs. We quantify this effect by
computing $\gamma_{\rm elec}
= \frac{1}{2}\int_{-\infty}^\infty\,\mrm{dz}\,\phi(z)n(z)$, which, as
can be seen in Fig.~\ref{fig4}c, decreases in an approximately
quadratic fashion as $\phi_0$ increases. By identifying $\alpha =
-\partial \gamma_{\mathrm{elec}} / \partial (\phi_{0}^2)$, we
reconstruct the potential-dependent contact angle, as shown in
Fig.~\ref{fig4}c. While the results of our microscopic theory are
broadly in line the macroscopic model (Eq. \ref{Eq:dielectrowetting})
some subtle differences are observed, especially at small
$\phi_0$. These appear to be correlated with significant changes in
the local compressibility near the interface, $\chi_\mu(z)
= \partial \rho(z) / \partial \mu$, as $\phi_0$ increases, which
indicates a suppression of density fluctuations. Such microscopic
details are lacking in the dielectric continuum model that underpins
Eq.~\ref{Eq:dielectrowetting}.

\section*{Conclusions}

Our findings establish EFGs as a powerful and versatile tool for
manipulating fluids. We have revealed their ability to structure
fluids, modulate phase transitions, and control capillary effects.
Crucially, we demonstrate that EFGs not only influence a fluid's
behavior in bulk, but also give rise to dielectrocapillarity, a new
phenomenon in which capillary condensation and criticality under
confinement can be finely tuned.
By placing this nanoscale
physics in direct correspondence with macroscopic
dielectrowetting experiments, our work provides a microscopic
foundation for the design of EFG-controlled wetting and adsorption
phenomena.  These discoveries are made possible by our development of
a multiscale approach that provides a first-principles description of
electromechanics \cite{bui2025hyperlmft}.

The effects uncovered in this work concern the equilibrium behavior of
dielectric liquids, and omit potentially important nonequilibrium
effects  such as pore entry and
exit \cite{Picard2021,Migacz2024,Giacomello2012}, electrokinetic
phenomena \cite{Shin2016,Shin2017,Marbach2019}, and controlled wetting
dynamics such as rate-dependent droplet spreading \cite{McHale2013}. Nonetheless, the implications of
our results for nonequilibrium behavior are potentially
far-reaching. A natural possible progression from this work is to
augment our first-principles framework for electromechanics with
dynamical extensions of cDFT \cite{Zimmermann2024}, opening a
promising route toward a microscopic understanding of how EFGs impact
non-equilibrium processes.
Moreover, the current framework naturally accommodates mixtures
of dielectric liquids -- in such cases, the excess intrinsic free
energy,
$\mcl{F}_{\mrm{intr}}^{\mrm{ex}}([\{\rho_{\nu}\},\beta\phi],T)$
acquires a functional dependence on the density fields of all species
present, with $\nu$ indexing each component. This generalization opens
the door to investigating more intricate phase behavior and interfacial
phenomena---including liquid--liquid phase separation driven by
EFGs \cite{Tsori2004,Tsori2007}.

The ability to reversibly control phase behavior and adsorption with
electric fields unlocks new avenues for manipulating
fluids across multiple length scales, from adaptive nanofluidic devices,
to tunable sorption in porous materials, to colloidal assembly.
At the nanoscale, nuclear magnetic resonance techniques
\cite{Liu2024} can directly validate these effects.
With strong EFGs experimentally accessible via
atomic force microscope  tips, optical tweezers, and 
patterned electrode configurations, our results lay the foundation for
future experimental exploration, paving the way for new strategies in
developing energy storage, selective separation, and responsive
fluidic technologies.

\section*{ Supporting information}
The supporting information includes the theory,
its practical implementation, details on the molecular models,
simulations, neural functional training procedure
and additional supporting results.

\section*{Data availability}
Data supporting the findings of this study will be openly
available upon publication of the manuscript \cite{zenodo}.

\section*{Code availability}
The code used to train the models in this study will be made publicly available upon publication of the manuscript \cite{zenodo}. Training data were generated using our in-house code for GCMC simulations \cite{githubgcmc} and the LAMMPS code for MD simulations~\cite{Thompson2022}.

\section*{Acknowledgments}

Via membership of the UK's HEC Materials
Chemistry Consortium funded by EPSRC
(EP/X035859), this work used the ARCHER2 UK
National Supercomputing Service. A.T.B. acknowledges funding from the Oppenheimer Fund and Peterhouse
College, University of Cambridge.  S.J.C. is a Royal Society
University Research Fellow (Grant No. URF\textbackslash
R1\textbackslash 211144) at Durham University.

\section*{Author contributions}
A.T.B.: Conceptualization (equal); Investigation (equal); Writing – original draft (equal); Writing – review \& editing (equal). S.J.C.: Conceptualization (equal); Investigation (equal); Writing – original draft (equal); Writing – review \& editing (equal)

{\small
\setlength{\parindent}{0pt}
\setlength{\parskip}{0.6em}
\section*{Methods}

\textbf{Reduced units.}
When reported, reduced units are defined by $T^* = k_{\rm B}T/\varepsilon$, $\rho^* = \rho\sigma^3$,
and $E^* = E\,\sigma^{3/2}\varepsilon^{-1/2}$, where $\sigma$ sets the molecular
length scale and $\varepsilon$ the energy scale.  Specifically for
each fluid: (i) SPC/E water, $\sigma = 3.166\,\text{\AA}$, $\varepsilon=0.65\,\mrm{kJ\,mol^{-1}}$;
the dipolar fluid, $\sigma = 3.024\,\text{\AA}$,
$\varepsilon=1.87\,\mrm{kJ\,mol^{-1}}$; and (iii) the electrolyte,
$\sigma=2.76\,\text{\AA}$, $\varepsilon=e^2/\sigma$, where $e$ is the
elementary charge.

\textbf{Hyperdensity functional theory.}
We employed the recently developed first-principles theory for
electromechanics
\cite{bui2025hyperlmft} based on hyperdensity functional theory
\cite{Sammuller2024hdft},
which provides an exact variational framework for the coupled
electromechanical response of fluids.  For a single-component fluid,
at specified $\mu$, $T$, planar external non-electrostatic potential
$V_{\mathrm{ext}}(z)$, and electrostatic potential $\phi(z)$, the
grand potential functional is
\begin{align*}
\varOmega([\rho,\beta\phi],T) = \mcl{F}^{\mrm{id}}_{\mrm{intr}}([\rho],T) &
  + \mcl{F}^{\mrm{ex}}_{\mrm{intr}}([\rho, \beta\phi],T) \nonumber\\
  & + \int\!\!\mrm{d}z\, \rho(z) [ V_{\rm ext}(z)- \mu],
\end{align*}
where the ideal intrinsic Helmholtz free energy is $\mcl{F}_{\rm
intr}^{\rm
id}([\rho],T)=k_{\mrm{B}}T\!\!\int\!\mrm{d}z\,\rho(z)[\ln \zeta^{-1}\Lambda^3\rho(z)-1]$,
with $\Lambda$ denoting the thermal de Broglie wavelength and
$\zeta$ denotes a partition function accounting for intramolecular
molecular degrees of freedom of each fluid particle.

The equilibrium number and charge density profiles are obtained by
solving the Euler--Lagrange (EL) equation
\[ 
\rho_{\rm eq}(z) = \frac{\zeta}{\Lambda^3}\exp\!\left[-\beta V_{\rm ext}(z) + \beta\mu + c^{(1)}(z;[\rho_{\rm eq},\beta\phi],T)\right],
\]
and evaluating
\[
n_{\rm eq}(z) = n^{(1)}(z;[\rho_{\rm eq},\beta\phi],T),
\]
with the one-body direct correlation functional $c^{(1)}$ and charge
density functional $n^{(1)}$ defined as first functional
derivatives of the excess intrinsic Helmholtz free energy functional
\[ c^{(1)}(z;
[\rho,\beta\phi], T) =
-\frac{\delta \beta \mcl{F}_{\mrm{intr}}^{\mrm{ex}}([\rho, \beta\phi],T)}{\delta \rho(z)},
\]
\[
n^{(1)}(z;[\rho,\beta\phi],T) =
\frac{\delta \mcl{F}^{\mrm{ex}}_{\rm intr}([\rho,\beta\phi],T)}{\delta\beta\phi(z)}.
\]

For fluids whose intermolecular interactions are long-ranged, such as
ionic and dielectric fluids, the functional dependence of
$c^{(1)}$ and $n^{(1)}$ on $\rho$ and $\phi$ is, in general,
non-local. Note that, as we only discuss these functionals 
evaluated at equilibrium in this work, we drop the ``eq'' subscript for notational convenience.

In order to learn these functional mappings, we applied
the techniques introduced in Refs.~\cite{bui2025hyperlmft,
bui2024learningclassicaldensityfunctionals} where the local
functionals of short-ranged mimic fluids,
$c_{\mathrm{R}}^{(1)}(z;[\rho,\beta\phi],T)$ and
$n_{\mathrm{R}}^{(1)}(z;[\rho,\beta\phi],T)$, were learned from
molecular simulation data using the neural functional
method \cite{Sammuller2023}. A SR mimic fluid is defined as a system
whose Coulomb interactions are replaced with
\[
\frac{1}{r} \rightarrow \frac{\mrm{erfc}(\kappa r)}{r}.
\]
For SPC/E water and the dipolar fluid, we used  $\kappa^{-1}=4.5\,\mrm{\AA}$.
For the electrolyte, we used the functionals trained in
Ref.~\onlinecite{bui2024learningclassicaldensityfunctionals}, with
$\kappa^{-1}=5.0\,\mrm{\AA}$. The effects of long-ranged
interactions are accounted for in a well-controlled mean field fashion
using local molecular field theory (LMFT).  Implementation details and
full derivations are provided in Ref.~\cite{bui2025hyperlmft} and
Section S1 of the SI.  We also validated the theory against computer
simulations in Section S3 of the SI.

\textbf{Generation of training data.} 
For the electrolyte, we employed the functional reported in
Ref.~\cite{bui2024learningclassicaldensityfunctionals} based upon
the restricted primitive model.  For SPC/E water and the dipolar
fluid, we generated training data using a combination of grand
canonical Monte Carlo (GCMC) and molecular dynamics (MD) simulations
to construct local reference functionals
$c_{\mathrm{R}}^{(1)}(z;[\rho,\beta\phi],T)$ and
$n_{\mathrm{R}}^{(1)}(z;[\rho,\beta\phi],T)$.  For each fluid, we
sampled $\sim$2000 randomized external conditions spanning both
subcritical and supercritical regimes.  Random inhomogeneous
electrostatic potentials of the form $\phi(z) = \phi_0 \cos(2\pi k z)$
were applied.  In some cases, planar walls of the form of a 9-3
Lennard--Jones potential were also included.  For each condition, GCMC
(performed with our own code \cite{githubgcmc}) determined the mean
particle number, which was then used to initialize MD simulations in
the NVT ensemble with LAMMPS~\cite{Thompson2022}, from which the
number and charge density profiles are sampled.  For water, we defined
the molecular center for sampling $\rho(z)$ at the oxygen site, and
for the dipolar fluid at the midpoint between the two opposite
charges. The charge density was computed as
$n(z) = \sum_i\sum^\prime_\alpha q_\alpha\langle\delta(z-z_{\alpha})\rangle$,
where the outer sum is over all molecules, and the inner primed sum is
over all sites belonging to molecule $i$, with $z_{\alpha}$ and
$q_{\alpha}$ indicating the $z$ coordinate and charge, respectively,
of site $\alpha$. This hybrid scheme reproduces the accuracy of pure
GCMC sampling while significantly accelerating convergence of
inhomogeneous structures.  The total computation time for the
generation of the entire dataset is on the order of $\sim 10^5$ CPU
hours.

\textbf{Training neural functionals.} 
For both SPC/E water and the dipolar fluid, we train two neural
networks to represent $c^{(1)}_{\mrm{R}}([\rho,\beta\phi],T)$ and
$n^{(1)}_{\mrm{R}}([\rho,\beta\phi],T)$ following the local learning
strategy \cite{Sammuller2023}.  The machine learning routine was
implemented in Keras/Tensorflow \cite{Chollet2017}.  Inputs consisted
of local density and electrostatic potential in a sliding spatial
window of size $10\,\mrm{\AA}$ from the center of the position of
interest.  To effectively learn spatial variations, the model
internally computes the gradient of $\beta\phi(z)$ using a central
difference scheme.  In addition to these spatially varying inputs, a
separate input node encodes $T$ as a scalar.  The full architecture
details are provided in Figs.~S5 and S6.  The dataset was split
roughly in a 3:1:1 ratio for training, validation, and test sets.
Models were trained for 200 epochs with a batch size of 256, using an
exponentially decaying learning rate starting at 0.001, achieving
errors comparable to the estimated simulation noise.  The training of
the neural networks was done on a GPU (\texttt{NVIDIA GeForce RTX
3060}) in a few hours. We also verified that the trained functional
can recover standard simulation results including bulk equation of
state (Fig.~S7) the binodal at zero applied field (Fig.~S8) as well as
number density and charge density response (Fig.~S9).

\textbf{Using the neural functionals.}  
 Evaluating the trained neural
functionals is fast ($\sim$ milliseconds) and can be performed on a CPU or GPU.
Combining with LMFT give us $c^{(1)}([\rho,\beta\phi],T)$ and 
$n^{(1)}([\rho,\beta\phi],T)$. 
The EL  equation  is solved self-consistently with a 
mixed Picard iteration scheme, which typically converges within minutes. To determine the liquid--vapor 
coexistence line at zero electric field, we calculate isotherms of the
chemical potential as a function of the bulk density $\rho_{\rm b}$
from  
\[
    \beta\mu = \ln(\Lambda^3 \rho_{\rm b}) - c^{(1)}([\rho_{\rm b},\beta\phi=0],T),
\]
and perform a Maxwell construction to find the coexisting liquid
and vapor densities at the binodal.  For inhomogeneous systems, i.e,
as a result of an applied non-uniform electric field, we set the
chemical potential and temperature and find the inhomogeneous
solutions by solving the EL equation. The isotherms of the chemical
potential, now as a function of the mean density,
$\overline{\rho}=L^{-1}\int^L_0\!\mrm{d}z\,\rho(z)$ where $L$ is the
total length of the domain over which $\rho(z)$ is defined, now have
distinct jumps when the system undergoes phase separation, giving
liquid $\rho_\mrm{l}$ and vapor $\rho_\mrm{v}$ solutions.  To
distinguish stable solutions from metastable solutions, we also
calculate the grand potential $\Omega_{\phi}
= \varOmega([\rho,\beta\phi],T)$ at a fixed $\beta\phi$ where the
excess free energy term can be evaluated via functional line
integration \cite{Sammuller2023}.  Performing this procedure for
$\beta\phi = 0$ gives results consistent with those obtained by
Maxwell construction. Note that the pressure of the bulk fluid at zero
field is given by $-PV =\beta\Omega_{0}$.  To investigate capillary
condensation in slit pores, adsorption isotherms are obtained from the
mean density $ \overline{\rho} = H^{-1}\int_0^H\!\!dz\,\rho(z)$ where
$H$ is the slit height.  Hysteresis loops are mapped by seeding
different initial guesses when solving the EL equation.}

In the
dielectrowetting calculations, the electrostatic free energy per
contact area is given as
\[
\gamma_{\mrm{elec}}=\frac{1}{2}\int^{+\infty}_{-\infty}\!\!\mrm{d}z\,\phi(z) n(z),
\]
where the factor of a half comes from there being two symmetric confining walls. 
To construct the dependence of $\theta$ on $\phi_0$, we obtain $\alpha = -\partial \gamma_{\mrm{elec}} / \partial (\phi_0^2)$  and calculate
$\gamma_{\mrm{lv}}$ from a direct coexistence molecular dynamics simulation. We estimate $\theta_0 = 160^\circ$ 
based on the maximum value of the local compressibility at $\phi_0 = 0$ and  comparing to Ref.~\cite{Evans2015}; 
while more detailed calculations can provide a more accurate estimate, this value is consistent with the solvophobic 
nature of the slit, and is sufficient for the purposes of demonstrating the increased wetting with $\phi_0$ shown in 
Fig.~4.

\bibliography{references}

\clearpage \includepdf[pages=1]{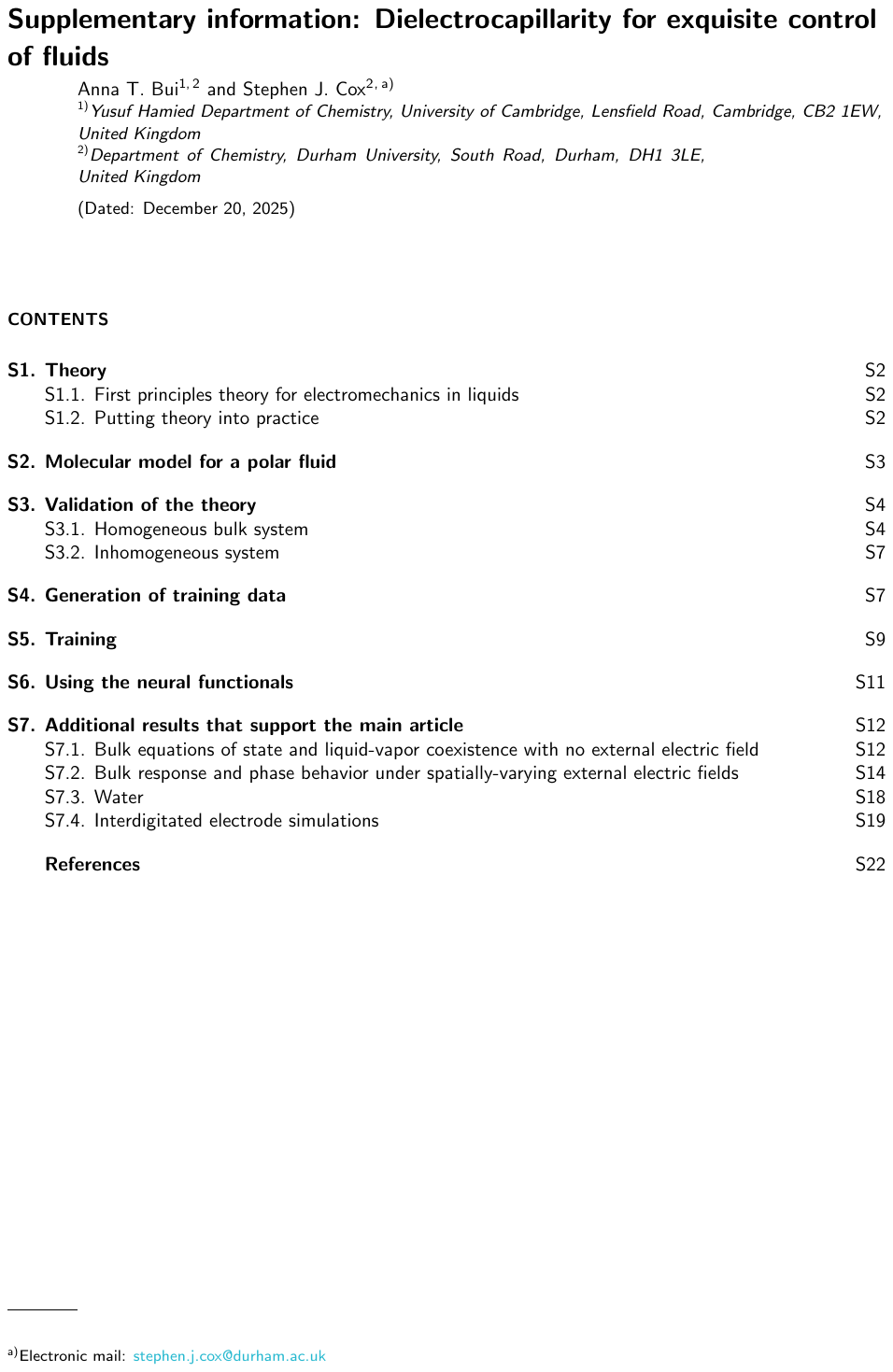}
\clearpage \includepdf[pages=2]{si.pdf}
\clearpage \includepdf[pages=3]{si.pdf}
\clearpage \includepdf[pages=4]{si.pdf}
\clearpage \includepdf[pages=5]{si.pdf}
\clearpage \includepdf[pages=6]{si.pdf}
\clearpage \includepdf[pages=7]{si.pdf}
\clearpage \includepdf[pages=8]{si.pdf}
\clearpage \includepdf[pages=9]{si.pdf}
\clearpage \includepdf[pages=10]{si.pdf}
\clearpage \includepdf[pages=11]{si.pdf}
\clearpage \includepdf[pages=12]{si.pdf}
\clearpage \includepdf[pages=13]{si.pdf}
\clearpage \includepdf[pages=14]{si.pdf}
\clearpage \includepdf[pages=15]{si.pdf}
\clearpage \includepdf[pages=16]{si.pdf}
\clearpage \includepdf[pages=17]{si.pdf}
\clearpage \includepdf[pages=18]{si.pdf}
\clearpage \includepdf[pages=19]{si.pdf}
\clearpage \includepdf[pages=20]{si.pdf}
\clearpage \includepdf[pages=21]{si.pdf}
\clearpage \includepdf[pages=22]{si.pdf}
\clearpage \includepdf[pages=23]{si.pdf}

\end{document}